%% file: main.tex
\documentclass{article}
\usepackage{arxiv}
\usepackage[utf8]{inputenc}
\usepackage[T1]{fontenc}
\usepackage{iftex}

\usepackage[square,numbers,sort&compress]{natbib}
\usepackage{hyperref}
\hypersetup{colorlinks=true,urlcolor=blue,linkcolor=black,citecolor=black}
\usepackage{url}
\usepackage{booktabs}
\usepackage{amsfonts}
\usepackage{nicefrac}
\usepackage{microtype}
\usepackage{lipsum}
\usepackage{graphicx}
\usepackage{threeparttable}
\usepackage{doi}
\usepackage{multirow}
\usepackage{multicol}
\usepackage{longtable}
\usepackage{xcolor}
\usepackage{subcaption}
\usepackage[normalem]{ulem}
\usepackage{amsmath,amsfonts,bm}
\usepackage{amssymb,amsthm}
\usepackage{enumitem}
\usepackage{adjustbox}
\usepackage{blindtext}
\usepackage{float}
\usepackage{array}
\usepackage{makecell}
\usepackage[perpage]{footmisc}
\usepackage{wrapfig}
\usepackage{tcolorbox}
\tcbuselibrary{breakable, skins}
\usepackage{tikz}

\ifPDFTeX
  \usepackage{CJKutf8}
  \AtBeginDocument{\begin{CJK*}{UTF8}{gbsn}}
  \AtEndDocument{\end{CJK*}}
  \newcommand{\zh}[1]{{#1}}
\else
  \usepackage{xeCJK}
  \newcommand{\zh}[1]{{#1}}
\fi

\definecolor{casebg}{RGB}{236, 241, 245}
\definecolor{thinkgray}{RGB}{80, 80, 80}

\newtcolorbox{CaseStudyBox}{
  colback=casebg,
  colframe=black,
  boxrule=1.2pt,
  arc=4mm,
  boxsep=8pt,
  left=10pt, right=10pt, top=10pt, bottom=10pt,
  breakable,
  enhanced,
  parbox=false,
}

\usepackage[capitalize]{cleveref}
\crefname{section}{\S}{\S\S}
\Crefname{section}{Section}{Sections}
\crefname{table}{Table}{Tables}
\Crefname{table}{Table}{Tables}
\crefname{figure}{Figure}{Figures}
\Crefname{figure}{Figure}{Figures}
\crefname{algorithm}{Algorithm}{Algorithms}
\Crefname{algorithm}{Algorithm}{Algorithms}

\setlist[itemize,1]{leftmargin=18pt, itemsep=2pt, topsep=3pt, parsep=2pt}
\setlist[enumerate,1]{leftmargin=18pt, itemsep=2pt, topsep=3pt, parsep=2pt}

\renewcommand{\arraystretch}{1.18}
\makeatletter
\renewenvironment{table}[1][tbp]
  {\@float{table}[#1]}
  {\end@float}
\makeatother
\title{\LARGE
\raisebox{-0.18\height}{\includegraphics[width=0.045\textwidth]{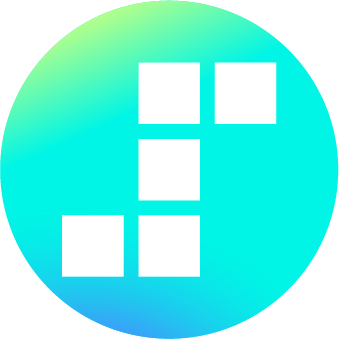}} %
DuplexSLA: A Full-Duplex Spoken Language Model with Synchronized Speech, Language, and Action
}

\author{\parbox{\textwidth}{\normalsize\centering
\mbox{Haoyang Zhang\textsuperscript{1,2,3,*}},
\mbox{Jun Chen\textsuperscript{1,*}},
\mbox{Donghang Wu\textsuperscript{3}},
\mbox{Yuxin Li\textsuperscript{1,3}}\\
\mbox{Yuxin Zhang\textsuperscript{1,4}},
\mbox{Xiangyu Tony Zhang\textsuperscript{1,5}},
\mbox{Che Liu\textsuperscript{1,6}},
\mbox{Qingjian Lin\textsuperscript{1}}\\
\mbox{Yizhou Peng\textsuperscript{3}},
\mbox{Hexin Liu\textsuperscript{3}},
\mbox{Eng Siong Chng\textsuperscript{3}},
\mbox{Chao Yan\textsuperscript{1}}\\
\mbox{Boyong Wu\textsuperscript{1}},
\mbox{Yechang Huang\textsuperscript{1}},
\mbox{Xuerui Yang\textsuperscript{1}},
\mbox{Fei Tian\textsuperscript{1,\textdagger}}\\[0.25em]
{\normalfont\small%
\mbox{\textsuperscript{1}StepFun}\quad
\mbox{\textsuperscript{2}Peking University}\quad
\mbox{\textsuperscript{3}Nanyang Technological University}\\
\mbox{\textsuperscript{4}Shanghai Jiao Tong University}\quad
\mbox{\textsuperscript{5}University of New South Wales}\quad
\mbox{\textsuperscript{6}Imperial College London}\\[0.1em]
\textsuperscript{*}Equal contribution.\quad
\textsuperscript{\textdagger}Corresponding author.}}}

\renewcommand{\headeright}{}
\renewcommand{\undertitle}{}
\renewcommand{\shorttitle}{DuplexSLA}

\begin{document}
\large

\maketitle

\input{content/abstract}
\input{content/introduction}
\input{content/architecture}
\input{content/data}
\input{content/post_train}
\input{content/evaluation}
\input{content/conclusion}

\input{content/Contributors}

\setlength{\bibsep}{0.5\baselineskip}
\bibliography{references}

\clearpage
\appendix
\begin{center}
{\LARGE\bfseries Appendix}
\end{center}
\vspace{1em}
\input{content/appendix}

\end{document}

%% file: content/abstract.tex
\begin{abstract}
Recent advances in spoken dialogue language models have shifted from turn-based to full-duplex designs, where the model continuously listens to the user while generating responses. However, existing duplex backbones still lack a native channel for in-conversation planning and tool calling, leaving real-time agentic behaviour either tied to turn boundaries or relegated to an external cascade. We propose \textbf{DuplexSLA}, a native full-duplex \emph{Speech--Language--Action} foundation model that decodes assistant audio together with a structured action stream on a shared \(160\) ms chunk timeline. DuplexSLA is built on a dual-stream three-channel formulation -- a continuous user audio channel, a discrete assistant audio channel, and a rate-limited textual action channel -- all decoded jointly by a single backbone, so that listening, speaking, planning, and tool calling unfold on one shared clock. Two capabilities define the model: (1) semantic-driven turn-taking control, where interruption, pause, and backchannel are handled inside the same backbone instead of by an external semantic VAD; and (2) in-conversation planning and tool calling, where planning text and structured tool calls are emitted on the action channel without halting assistant audio, so that multi-action and backchannel-triggered tool use are interleaved with ongoing speech. To evaluate these capabilities together, we further construct \textbf{DuplexSLA-Bench}, a duplex benchmark covering pause, interrupt, and backchannel turn-taking together with three styles of in-conversation tool calling. Our project page, interactive demos, and the DuplexSLA-Bench evaluation suite are publicly available at \url{https://github.com/hyzhang24/DuplexSLA}.
\end{abstract}

%% file: content/introduction.tex
\section{Introduction}
\label{sec:intro}

Natural conversation is not a strict alternation between two speakers. Listeners begin to plan a response before a turn ends, tolerate hesitations, send short feedback such as ``mm-hmm'' without taking the floor, and recover quickly from accidental overlap. They also \emph{act} while talking: opening an application or triggering a control on the same conversational clock as their words. A spoken assistant that lacks any of these behaviours feels rigid no matter how strong its underlying language model is.

Most deployed speech agents still rely on a turn-based pipeline of VAD, ASR, LLM, and TTS, which encounters two structural problems in duplex spoken interaction. First, an energy-based VAD cannot distinguish an end-of-turn silence from a hesitation pause, a brief backchannel, or a real interruption; bolting an external semantic VAD on top recovers part of this nuance but adds latency and still does not see the assistant's internal state. We find that integrating these decisions into a native full-duplex backbone is much more effective than externally attaching another semantic VAD: a sufficiently large duplex model with adequate data absorbs pause, backchannel, and interruption phenomena into its core conversational competence, rather than paying the latency of an extra detector chain. Second, tool calling does not fit cleanly into a turn-based loop. Emitting tool calls before the assistant speaks adds wall-clock delay; emitting them after the assistant finishes delays the side-effect by a full turn; and emitting them mid-utterance through the same channel that drives speech tends to break the spoken response. What is missing is a model that can listen, speak, think, and act on \emph{one} synchronized timeline. Recent work has therefore moved toward native full-duplex speech models that learn to listen and speak inside a single backbone~\citep{wang2024fullduplex,veluri2024synchronous,defossez2024moshi,wang2024freezeomni,yu2024salmonnomni,hu2025salmduplex,wang2026covoaudio,roy2026personaplex,xie2024miniomni,xie2024miniomni2,fang2024llamaomni,zhang2024omniflatten,nguyen2025spiritlm,yu2025voila} and toward chunk-aligned reasoning on the same timeline as audio~\citep{wu2025chronological,wu2025mind,wu2026silent}, with audio-aware foundation models~\citep{team2024qwen2,chu2024qwen2,huang2025stepaudio,wu2025step,tian2025stepaudior1,zhang2026stepaudior15,openai2024gpt4o,xu2025qwen3,zeng2024glm4voice,fu2025vita15,li2025vitaaudio,minicpmo2025,zhang2025mamba,liu2026code,liu2026boosting} and existing duplex and spoken-language benchmarks~\citep{lin2025fullduplexbench,zhang2025wildspeech,wang2025mmsu,sakshi2024mmau,deng2025multibench,arora2025talkingturns,chen2024voicebench,yang2024airbench,ao2024sdeval,yan2025urobench,liu2025vocalbench} as supporting context. We focus on a combination that existing duplex backbones and benchmarks do not jointly stress: semantic-driven turn-taking control \emph{plus} in-conversation tool calling.

We propose \textbf{DuplexSLA}, a native full-duplex foundation model with a dual-stream three-channel formulation. The model continuously consumes user audio, decodes assistant audio, and emits textual action tokens in lockstep on a shared \(160\) ms chunk grid. Each chunk carries (1) two \(80\) ms causal user audio features, (2) four \(40\) ms discrete assistant audio tokens preceded by a text anchor (a TA4 layout), and (3) up to ten action text tokens that may contain delayed transcript text, planning text, control labels (\texttt{interrupt}, \texttt{backchannel}, \texttt{response}), and structured tool calls. The same backbone autoregressively predicts the assistant TA4 stream and the action stream, while the user audio side is kept causal and is never produced by the model. The two highlight capabilities of DuplexSLA are: (1) semantic-driven native interruption, pause, and backchannel, where chunk-level turn-taking decisions live on the action channel of the same backbone that drives assistant speech, removing the external semantic VAD; and (2) in-conversation planning and tool calling, where planning text and JSON-style tool calls are emitted on the action channel while the assistant TA4 channel keeps producing speech, all under a strict per-chunk action-token budget (\cref{sec:arch}).

The action channel is what makes the third letter of \emph{Speech--Language--Action} non-trivial: it gives every action object and VAD-like decision (\texttt{interrupt}, \texttt{backchannel}, \texttt{response}) a dedicated, time-stamped textual lane co-decoded with assistant audio, instead of either competing for slots in the assistant text channel or being relegated to a post-hoc cascade.

The contributions of this report are as follows.
\begin{itemize}[itemsep=2pt, topsep=2pt, parsep=0pt]
    \item We propose \textbf{DuplexSLA} (\cref{sec:arch}), a native full-duplex foundation model that co-decodes assistant audio and a structured action stream on a shared chunk timeline.
    \item We construct \textbf{DuplexSLA-Bench} (\cref{sec:eval}), a \(2{,}100\)-case duplex evaluation suite with a timing-aware tool-call protocol, on which DuplexSLA reaches sub-second latency while remaining competitive on tool-call accuracy.
\end{itemize}

%% file: content/architecture.tex
\section{Model Architecture}
\label{sec:arch}

\subsection{Dual-stream three-channel formulation}

DuplexSLA models a spoken interaction as a sequence of fixed-size chunks. Let \(\Delta = 160\) ms be the chunk size on the conversational clock, with time discretized as \(c = \lfloor t / \Delta \rfloor\). At each chunk \(c\), the model receives one user audio segment and one assistant audio segment, and produces two outputs: an assistant audio segment and an action segment, both indexed by \(c\).

\begin{figure*}[t]
\centering
\includegraphics[width=\textwidth]{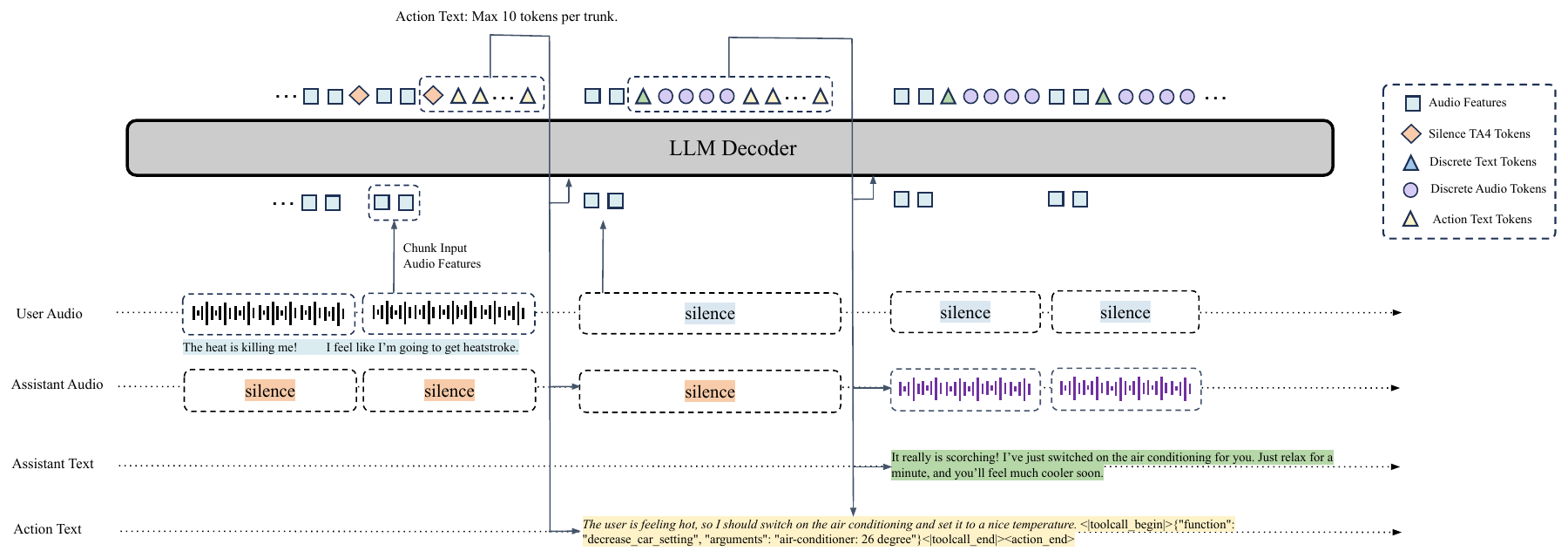}
\caption{\textbf{DuplexSLA chunk-level architecture.} Each chunk is \(160\) ms. The user channel contributes \(2\) causal audio features (\(80\) ms each); the assistant channel contributes a TA4 unit (one text anchor and \(4\) discrete audio tokens at \(40\) ms each); the action channel emits up to \(10\) text tokens that may be delayed transcript text, planning text, or tool calls. The same backbone autoregressively predicts the assistant TA4 and the action text.}
\label{fig:arch_chunk}
\end{figure*}

We organize each chunk into three channels:
\begin{itemize}
    \item \textbf{User Channel}: a continuous user audio feature sequence. Each chunk contributes \(2\) causal features at an \(80\) ms stride.
    \item \textbf{Assistant Channel}: a discrete assistant speech sequence in \emph{TA4} layout. Each chunk emits one text anchor token \(T\) followed by \(4\) audio tokens \(A\) at a \(40\) ms stride.
    \item \textbf{Action Channel}: a textual stream that may contain delayed transcript text, planning text, interaction-control labels, or structured tool calls. The action stream is rate-limited (see \cref{sec:budget}).
\end{itemize}
We refer to this design as the dual-stream three-channel formulation: there are two physical audio streams (user and assistant) on the conversational clock, but three semantic channels on the model interface, because the action channel is text-only and lives on top of the assistant timeline. As illustrated in \cref{fig:arch_chunk}, the LLM decoder consumes user audio features together with previously generated assistant audio and action text, and produces the next chunk's assistant audio plus action text in a single autoregressive step.

\subsection{Per-chunk serialization}

Within a chunk, the three channels are interleaved into a single token stream consumed by the LLM backbone. The serialization is:
\begin{tcolorbox}[
  colback=gray!4, colframe=black!55, boxrule=0.4pt, arc=2pt,
  boxsep=3pt, left=10pt, right=10pt, top=4pt, bottom=4pt
]
\centering
\small
\renewcommand{\arraystretch}{1.15}
\begin{tabular}{@{}l@{\quad}c@{\quad}l@{}}
\texttt{<|user\_audio\_begin|>}      & \(U\ U\)          & \texttt{<|user\_audio\_end|>}      \\
\texttt{<|assistant\_audio\_begin|>} & \(T\ A\ A\ A\ A\) & \texttt{<|assistant\_audio\_end|>} \\
\multicolumn{2}{@{}l@{}}{\(\langle\text{action text}\rangle\)}            & \texttt{<|action\_end|>}           \\
\end{tabular}
\end{tcolorbox}
The user audio segment is encoded by a causal speech front end (so the streaming property is preserved), while the assistant TA4 unit and the action text are the parts the model must \emph{produce}. Whenever the chunk has nothing to say, \(T\) is predicted as a special anchor token (\texttt{<vad\_silence>} or \texttt{<tts\_pad>}) and the four \(A\) tokens are predicted as the corresponding silence audio codes. The \texttt{<|action\_end|>} marker terminates the chunk regardless of whether any action text was emitted, which keeps every chunk strictly aligned to the \(160\) ms clock.

For the action channel, we use a small set of structured markers in addition to free text:
\begin{itemize}
    \item planning text: a short rationale fragment;
    \item turn-taking labels: \texttt{response} / \texttt{interrupt} / \texttt{backchannel};
    \item tool call: a JSON body wrapped in \texttt{<|toolcall\_begin|>}\,$\ldots$\,\texttt{<|toolcall\_end|>} that names a function and its arguments.
\end{itemize}
The action segment for a chunk that carries planning text plus a single tool call therefore has the following abstract form:
\begin{tcolorbox}[
  colback=gray!4, colframe=black!55, boxrule=0.4pt, arc=2pt,
  boxsep=3pt, left=10pt, right=10pt, top=3pt, bottom=3pt,
  fontupper=\footnotesize
]
\texttt{planning<|toolcall\_begin|>\{"function":\,"function\_name",\,"arguments":\,"arguments"\}<|toolcall\_end|>}
\end{tcolorbox}
Here \texttt{planning} is a free-form rationale fragment, and the JSON body inside \texttt{<|toolcall\_begin|>}/\texttt{<|toolcall\_end|>} carries the function name and structured arguments. Both the action text and the assistant TA4 unit produced in chunk \(c\) are aligned to chunk \(c\) on the conversational clock, so a tool call emitted while the assistant is still speaking can be assigned a precise time stamp by reading the chunk index. Compared with backbones that work with two streams only (user and assistant), this explicit separation pushes planning and tool calling onto a dedicated, time-stamped channel without disturbing the assistant audio.

\subsection{Real-time decoding budget}
\label{sec:budget}

Real-time full-duplex interaction requires the per-chunk decoding cost of the model to fit inside one \(160\) ms chunk on the actual inference hardware. After the assistant TA4 unit is paid for, the autoregressive decoding throughput of a \(7\)B-scale backbone on mainstream inference accelerators leaves room for only a small number of action-channel tokens per chunk. We therefore cap the action channel at 10 text tokens per chunk, with a safe margin against the per-chunk wall-clock budget; tokens that do not fit spill into the action segments of the following chunks. This bound is a deployment budget, not an architectural constraint, and can be re-tuned per accelerator without retraining.

\subsection{Native interruption, pause and backchannel}

\begin{figure}[H]
\centering
\includegraphics[width=\textwidth]{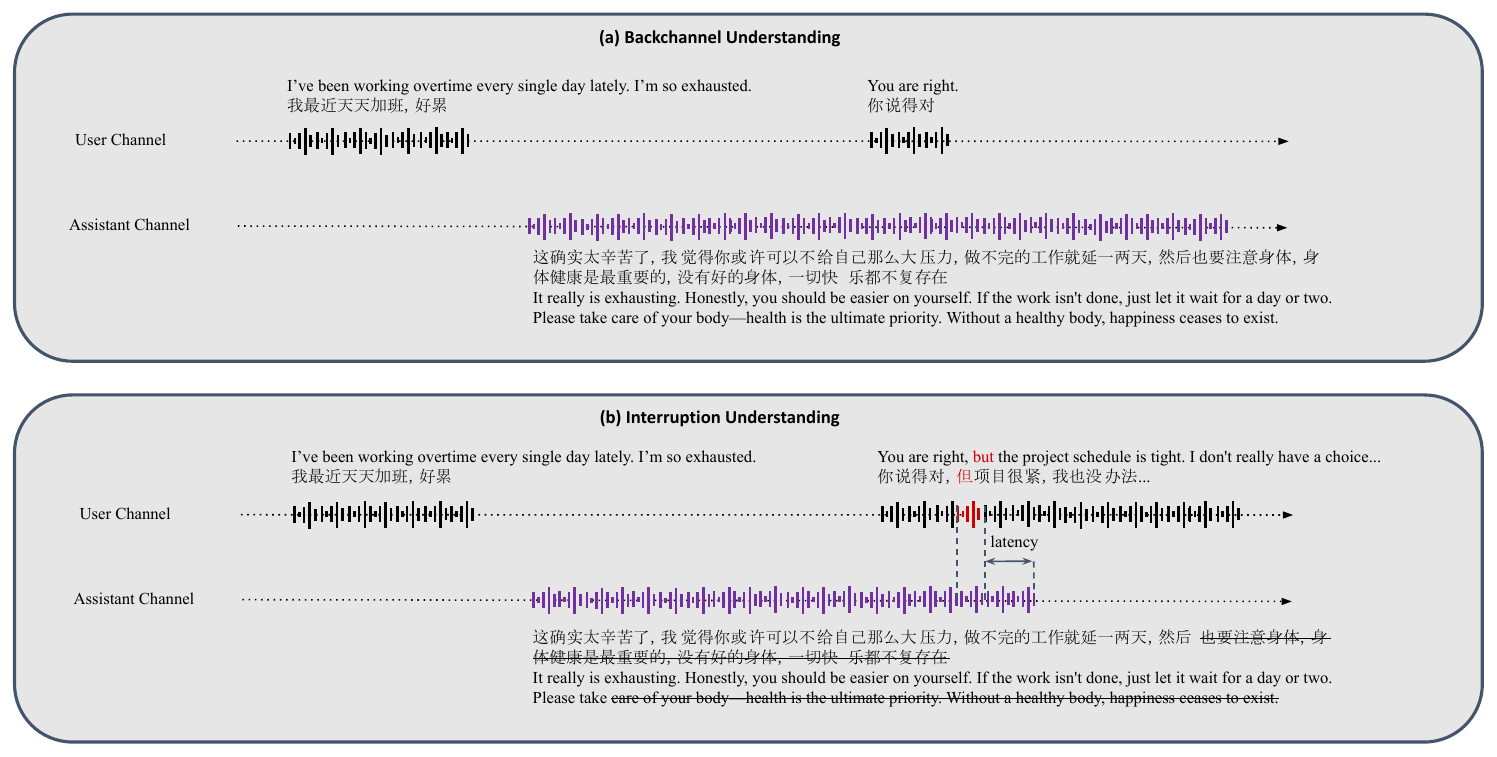}
\caption{\textbf{Native interaction-control behaviours.} (a) A short user backchannel (``You are right'') does not stop the assistant; the action channel emits a backchannel label while assistant speech keeps flowing. (b) When the user starts a real new thought (``You are right, but the project schedule is tight\ldots''), DuplexSLA emits an interrupt label and the assistant yields the floor within a small chunk-level latency.}
\label{fig:interrupt_backchannel}
\end{figure}

Because the action channel shares a backbone with assistant speech generation, control decisions can be derived from the same internal representation that drives the response. Three semantic phenomena therefore become intrinsic model behaviours rather than external rules:
\begin{itemize}
    \item \textbf{Pause}: the user holds their thought without ending the turn, and the assistant should remain silent. The action stream stays at \texttt{response}-style continue-listening labels, and the assistant TA4 keeps emitting silence anchors.
    \item \textbf{Interrupt}: the user starts a new thought while the assistant is speaking. Around the semantic interruption point, DuplexSLA emits an \texttt{interrupt} label on the action channel and switches the assistant TA4 to silence within a small number of chunks.
    \item \textbf{Backchannel}: the user produces a short feedback utterance (e.g., ``yes'', ``you are right'') without intending to take the floor. The action channel emits a \texttt{backchannel} label, and the assistant continues without resetting its current speech plan.
\end{itemize}
Crucially, all three decisions are made based on the model's internal semantic state, rather than on a separate VAD module. The benefit is concrete: a turn-based pipeline that bolts a semantic VAD on top still has to pay the latency of the external detector and is bounded by what that detector can express. With sufficient duplex data, a large native duplex model integrates these decisions directly into its token-level dynamics, as reflected in the timing numbers reported in \cref{sec:eval}. \cref{fig:interrupt_backchannel} illustrates the two intuitive cases: a backchannel that does not stop the assistant and an interruption that does.

\subsection{In-conversation planning and tool calling}

The action channel also carries planning text and structured tool calls, which is what keeps tool calling \emph{on-line} rather than a turn-final dump. Two patterns are particularly important.

\paragraph{Backchannel-triggered tool calling.}
A short user utterance that is topically unrelated to the current dialogue (e.g., ``play some Beatles songs'' uttered while the assistant is talking about something else) is treated as a backchannel: the action channel emits a planning fragment plus a tool call (\texttt{play\_music("The Beatles")}) without interrupting the assistant's spoken thread. The assistant audio thus stays coherent while the side-effect is dispatched.

\paragraph{Multi-action tool calling.}
A single user request can spawn several tool calls, e.g., raising the AC, playing music, and navigating to a restaurant. Because each tool call is anchored to its own chunk on the action channel, the calls are emitted in semantic order along the user's request, and the assistant audio runs in parallel with each call's planning text. This pattern is difficult to express cleanly in turn-based agents, where multi-tool plans either delay all speech until the tools resolve or break the spoken response.

\begin{figure}[H]
\centering
\includegraphics[width=\textwidth]{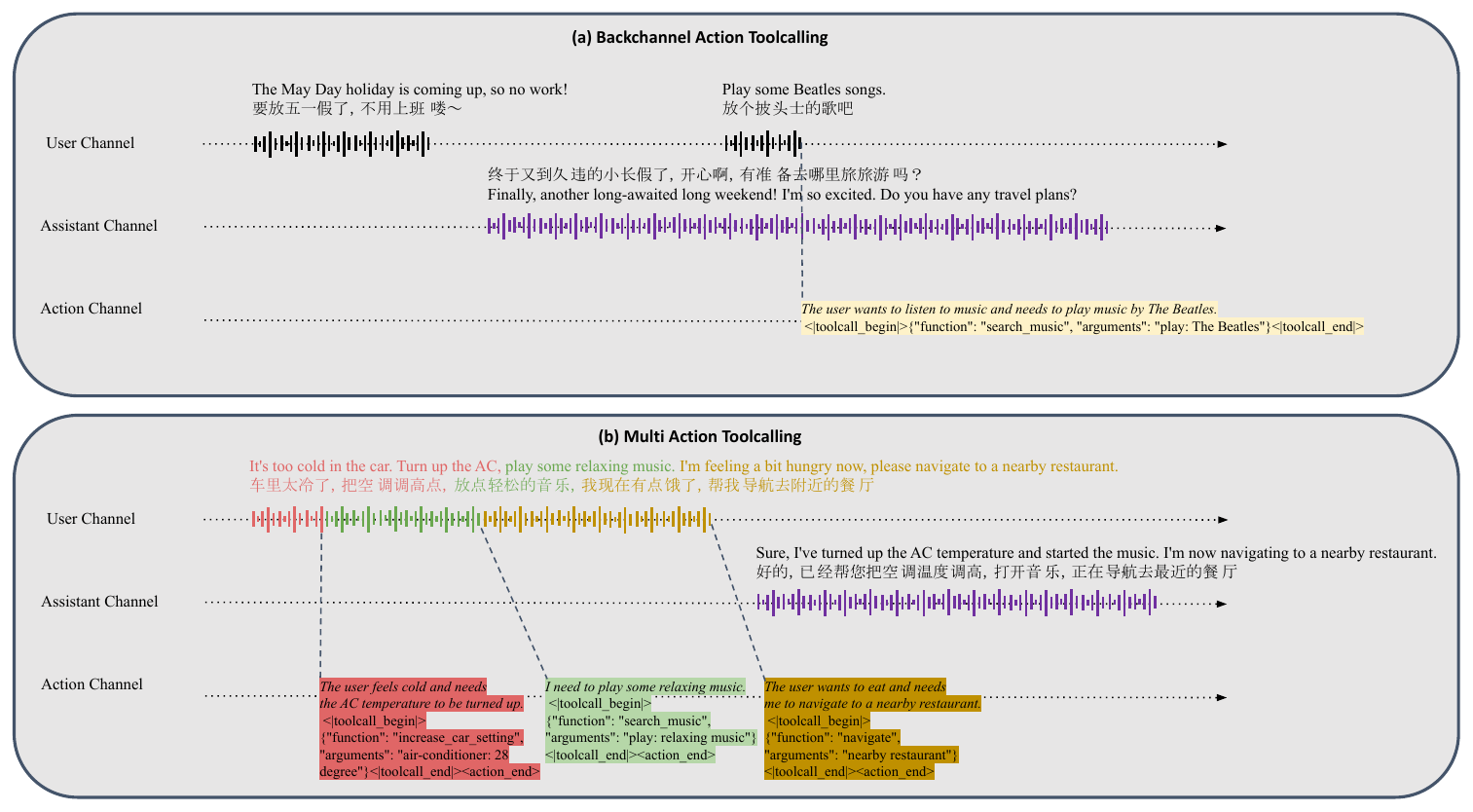}
\caption{\textbf{Planning + tool-call integration on the action channel.} (a) Backchannel-triggered tool call: a brief user utterance triggers \texttt{play\_music("The Beatles")} on the action channel while the assistant keeps speaking. (b) Multi-action tool calling: one user turn with three intents emits three time-aligned tool calls (AC, music, navigation) on the action channel; assistant speech and action emission run on the same chunk timeline.}
\label{fig:toolcall_capabilities}
\end{figure}

\cref{fig:toolcall_capabilities} shows both patterns. In the first row, the user issues a backchannel-style request and the action channel emits a tool call without disturbing the assistant. In the second row, a single user turn produces three time-aligned tool calls, each anchored to the relevant chunk on the action channel.

\subsection{Action channel design rationale}

The dedicated third channel is what makes the two highlight capabilities cheap to learn and to serve. Embedding planning and tool calls into the same channel as assistant text would force that channel to alternate between TA4 audio tokens and tool-call JSON, which breaks the smoothness of the assistant audio. A separate action channel that is itself synchronized to the chunk clock also gives every action object an unambiguous timestamp, which is needed both for downstream execution and for the latency-oriented evaluation in \cref{sec:eval}. The cost of the third channel is modest -- at most \(10\) tokens per chunk -- and is easily absorbed by the per-chunk decoding budget in \cref{sec:budget}.

A condensed view of the system is given in \cref{tab:model_config}.

\begin{table}[h]
\centering
\small
\begin{tabular}{p{0.30\linewidth}p{0.66\linewidth}}
\toprule
Design element & Current formulation \\
\midrule
Backbone scale & \(7\)B speech-LM, initialized from Step-Audio~2 mini~\citep{wu2025step} \\
Streaming clock & \(160\) ms conversational chunks \\
User audio granularity & \(2\) causal acoustic features per chunk (\(80\) ms each) \\
Assistant audio granularity & \(4\) discrete audio tokens per chunk (\(40\) ms each) \\
Per-chunk speech layout & TA4 (one text anchor + four audio tokens) \\
Action channel content & Delayed transcript text, planning text, turn-taking labels, tool calls \\
Per-chunk action token budget & \(\le 10\) tokens; overflow spills into next chunks \\
Tool-call schema & \(50\) cabin and smart-home function schemas, plus \(3\) interaction-control labels \\
Native duplex behaviours & Pause, interrupt, and backchannel without external semantic VAD \\
Online tool calling & Backchannel-triggered, single-action, multi-action \\
\bottomrule
\end{tabular}
\caption{System-level summary of DuplexSLA.}
\label{tab:model_config}
\end{table}

%% file: content/data.tex
\section{Data Construction}
\label{sec:data}

\begin{figure*}[t]
\centering
\includegraphics[width=\textwidth]{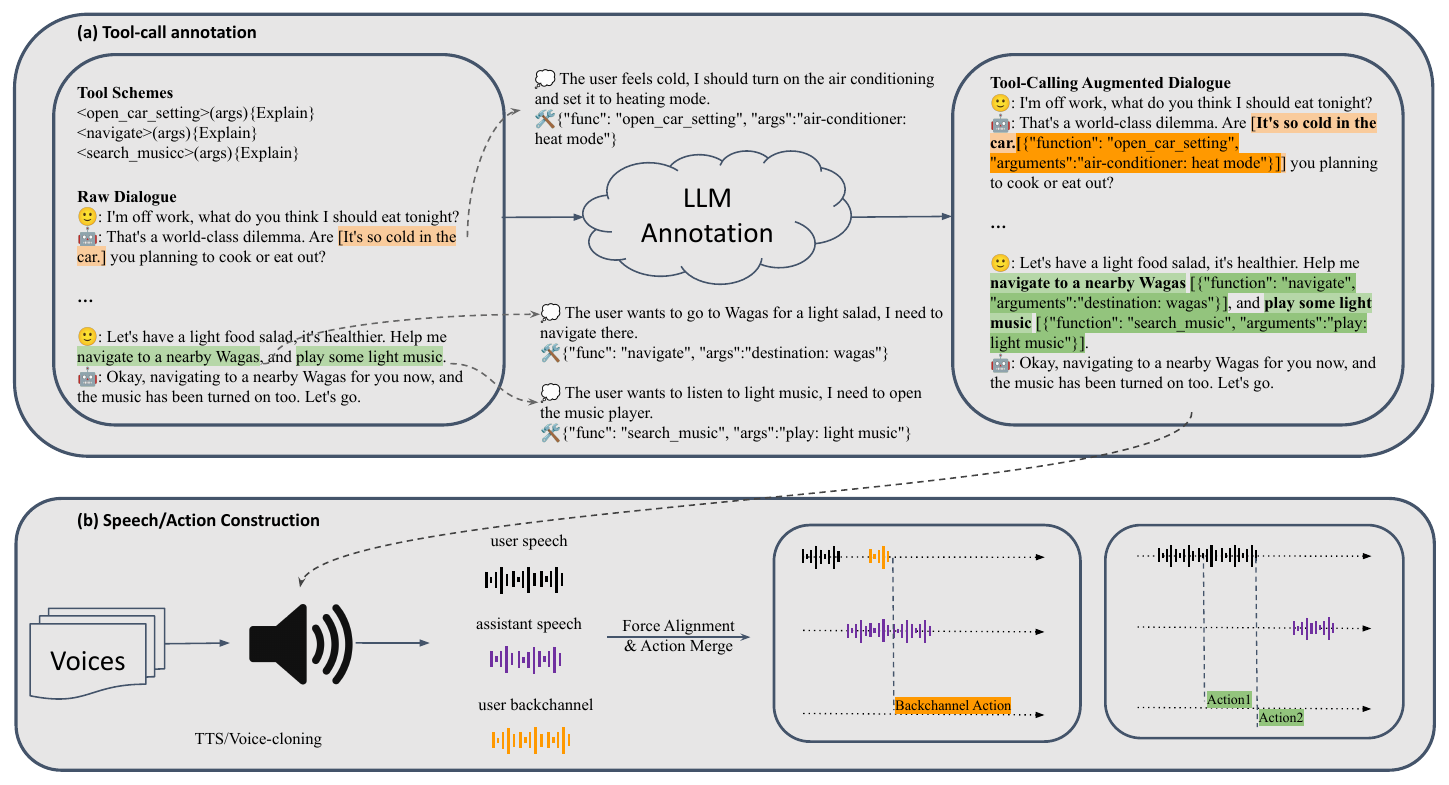}
\caption{\textbf{Data-construction pipeline.} (a) An LLM annotates each raw dialogue with tool-call objects (function name, arguments, planning text, semantic offset). (b) The user and assistant utterances are synthesized with TTS and voice cloning, force-aligned, time-merged, and the action-channel labels (backchannel, interrupt, planning, tool calls) are merged at the chunk grid.}
\label{fig:data_pipeline}
\end{figure*}

The chunked, dual-stream three-channel format described in \cref{sec:arch} does not match the format of conventional dialogue corpora, so building DuplexSLA required a dedicated data-construction effort. The goal of this section is to give a clear picture of \emph{the data form} fed to the model and \emph{the mixture proportions} used in training, rather than to enumerate every annotation detail. \cref{fig:data_pipeline} summarises the pipeline in two stages.

\subsection{Sample form}

Every training sample is a chunked dual-track session. The shared schema is as follows:
\begin{itemize}
    \item a task-conditioned system prompt (one of \texttt{dialogue}, \texttt{asr\_human}, \texttt{asr\_assistant}, \texttt{interrupt}, \texttt{backchannel}, \texttt{pause}, \texttt{toolcall});
    \item a user audio track and an assistant audio track aligned on the same conversational clock; the assistant audio is represented as discrete speech units (\(4\) per chunk in TA4 layout);
    \item an ordered list of action objects, each with a function name, optional planning text, optional structured arguments, and a semantic trigger offset that is snapped to a chunk index at training time;
    \item the assistant TA4 stream and the action stream are both supervised, while the user audio side is observed only.
\end{itemize}
The same schema covers all task families: they differ only in which channels carry information. Specifically, ASR families use the action channel for delayed transcript text, timing-control families use it for the \texttt{interrupt}, \texttt{backchannel}, and \texttt{response} labels, and tool-use families use it for planning text plus structured tool calls. A more detailed enumeration of the action vocabulary is given in \cref{app:vocab}, and full per-task chunk-by-chunk traces in \cref{app:serialization}.

\subsection{Mixture proportions}

\begin{figure*}[t]
\centering
\includegraphics[width=0.85\textwidth]{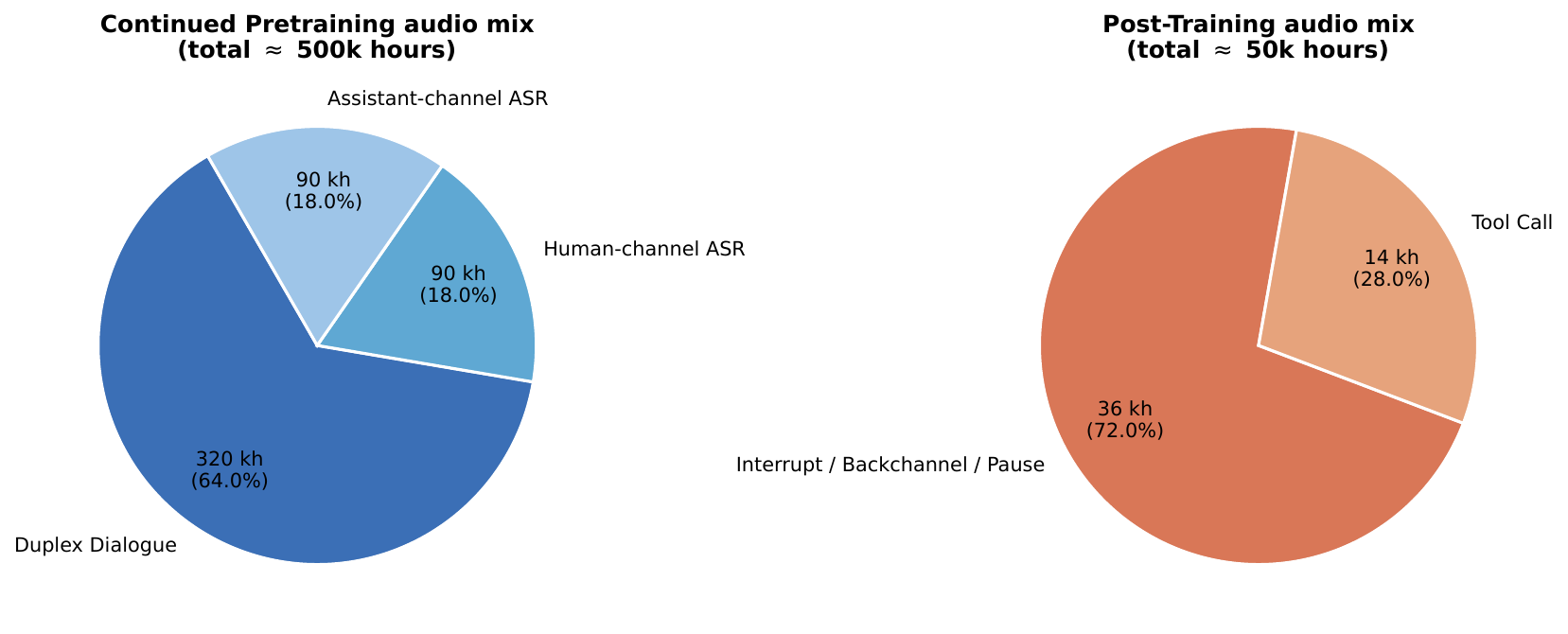}
\caption{Audio-data distribution across continued pretraining (left) and post-training (right). CPT is dominated by duplex dialogue, with substantial dual-side ASR supervision; post-training is dominated by interaction-control data, with a smaller but capability-critical tool-call slice.}
\label{fig:data_pie}
\end{figure*}

The corpus is a mixture of seven task families across two stages of training, jointly chosen so that the backbone preserves its world knowledge and language ability, internalises the chunked dual-stream three-channel format, and learns tight time alignment between audio and the action channel. Continued pretraining (CPT) is dominated by ordinary duplex dialogue, augmented with substantial dual-side ASR supervision and general text data. Post-training is much smaller in volume, but is concentrated on the timing-sensitive and action-emitting behaviours that DuplexSLA is built to deliver: turn-taking control cases (pause, interrupt, and backchannel), and three styles of tool-call data (single-action turn-taking, multi-action turn-taking, and backchannel-action). The proportion split is shown in \cref{fig:data_pie}, and the concrete scales are listed in \cref{tab:data_mix}.

\begin{table}[h]
\centering
\small
\begin{tabular}{p{0.55\linewidth}r}
\toprule
Data family & Scale \\
\midrule
\multicolumn{2}{@{}l}{\emph{Continued pretraining (\(\sim\)500\,k hours audio + \(\sim\)1.92\,M text samples).}} \\
Text                                  & \(\sim\)1.92\,M samples \\
Duplex dialogue                       & \(\sim\)320\,k hours \\
User-channel ASR                     & \(\sim\)90\,k hours \\
Assistant-channel ASR                 & \(\sim\)90\,k hours \\
\midrule
\multicolumn{2}{@{}l}{\emph{Post-training (\(\sim\)50\,k hours).}} \\
Interrupt + backchannel + pause       & \(\sim\)36\,k hours \\
Tool-call (BC-action, single-action, and multi-action) & \(\sim\)14\,k hours \\
\bottomrule
\end{tabular}
\caption{Training-data mixture across the two stages. CPT preserves language ability and teaches the dual-stream three-channel format; post-training concentrates on the capability-critical interaction-control and tool-call slices reported in \cref{sec:eval}.}
\label{tab:data_mix}
\end{table}

\subsection{Action objects on the timeline}
\label{sec:fifo}

Each action object is anchored to its semantic trigger time on the conversational clock. At training time, these offsets are snapped to chunks and emitted on the action channel of the corresponding chunk. Because every chunk has a hard \(\le\!10\)-token action-channel budget (\cref{sec:budget}), short bursts of actions cannot always fit into the chunk where they are triggered. The data-construction format therefore turns the action stream into a single FIFO queue keyed by trigger time, with two consistent rules:
\begin{itemize}[itemsep=2pt, topsep=3pt, parsep=0pt]
    \item \textbf{Within a chunk.} If two or more actions are triggered in the same chunk, they are serialized in trigger-time order (ties broken by id) and concatenated head-to-tail on that chunk's action channel. The chunk-terminating \texttt{<|action\_end|>} marker is emitted \emph{only} after the last queued action has fully closed, so an open \texttt{<|toolcall\_begin|>}\,$\ldots$\,\texttt{<|toolcall\_end|>} block is never split by an early \texttt{<|action\_end|>}.
    \item \textbf{Across chunks.} If an action's planning text plus tool-call body exceeds the per-chunk \(\le\!10\)-token budget, the surplus tokens spill into the action segments of the following chunks. Any later-triggered action waits in the queue until the in-flight action has fully drained, and then starts emitting from the next available chunk: it never preempts an earlier action, and never breaks an open \texttt{<|toolcall\_begin|>}\,$\ldots$\,\texttt{<|toolcall\_end|>} block.
\end{itemize}
Concretely, the within-chunk rule produces concatenated action segments of the schematic form below (shown with two queued actions before any chunk-level truncation is applied):
\begin{tcolorbox}[
  colback=gray!4, colframe=black!55, boxrule=0.4pt, arc=2pt,
  boxsep=3pt, left=8pt, right=8pt, top=4pt, bottom=4pt,
  fontupper=\footnotesize\ttfamily
]
planning\_1\\
<|toolcall\_begin|>\{"function":\,"function\_name\_1",\,"arguments":\,"arguments\_1"\}<|toolcall\_end|>\\
planning\_2\\
<|toolcall\_begin|>\{"function":\,"function\_name\_2",\,"arguments":\,"arguments\_2"\}<|toolcall\_end|>\\
<|action\_end|>
\end{tcolorbox}
Each \texttt{<|toolcall\_begin|>}/\texttt{<|toolcall\_end|>} block stays atomic; the chunk-terminating \texttt{<|action\_end|>} appears only after the very last action in the queue has been written out, and any tokens that overshoot the per-chunk budget are deferred to the following chunks under the across-chunk rule above.

Because the assistant TA4 channel has its own per-chunk token budget, this FIFO queue on the action channel never blocks assistant speech: while the queue is draining over several chunks, the TA4 stream keeps producing audio in lockstep. This is what allows the same backbone to learn ``what to do'' and ``when to do it'' from the same supervision target, and is also why the multi-action tool calls in \cref{fig:toolcall_capabilities}b are emitted in the order of the user's request without delaying the spoken response.

\subsection{Dual-side ASR is required for time alignment}

A subtle but important point is that the action channel learns \emph{both} a delayed user transcript and a delayed assistant transcript. Inside the assistant TA4 layout, the text anchor \(T\) is left-aligned within the chunk and does not carry exact timing. By forcing the action channel to emit the assistant transcript at the chunk where each character is actually being spoken, we explicitly tie assistant audio to action time. As a result, the model's internal time clock stays consistent across user audio, assistant audio, and action emission, which is what makes sub-second tool-call latency feasible.

%% file: content/post_train.tex
\section{Training Recipe}
\label{sec:training}

DuplexSLA is initialized from \textbf{Step-Audio~2 mini}~\citep{wu2025step}, a \(7\)B-scale audio language model. Training proceeds in two stages: continued pretraining (CPT) brings the backbone into the chunked dual-stream three-channel format, and post-training instils the four duplex-interaction behaviours and the three tool-call patterns evaluated in \cref{sec:eval}.

\subsection{Stage 1: continued pretraining}

The goal of CPT is to make the backbone fluent in the new serialization. The model has to learn three new things at once: (1) the chunk-level interleaving of user audio, assistant TA4, and action text; (2) the strict time alignment between assistant audio and action text via dual-side ASR; and (3) silence behaviours on both the assistant TA4 anchor (\texttt{<vad\_silence>}, \texttt{<tts\_pad>}) and the action channel.

The CPT mixture is dominated by audio data (\(\sim 500\)k hours in total), supplemented with general text data to preserve world knowledge and reasoning ability:
\begin{itemize}
    \item Duplex dialogue (\(\sim 320\)k h) supplies the base distribution of full-duplex spoken interaction.
    \item User- and assistant-channel ASR (\(2 \times 90\)k h) supply explicit time alignment between audio and action text.
    \item Text (\(\sim 1.92\)M samples) preserves language and reasoning ability while the speech format changes.
\end{itemize}
After CPT, the model becomes comfortable with the duplex serialization, but does not yet exhibit the targeted real-time interaction behaviours, especially when the user pauses, interrupts, or issues short backchannel feedback.

\subsection{Stage 2: capability-oriented post-training}

Post-training shifts the data distribution from generic duplex dialogue toward the behaviours we want to evaluate. The mixture is deliberately small (\(\sim 50\)k hours), but each slice is highly informative. Two families dominate (post-training block of \cref{tab:data_mix}):
\begin{itemize}
    \item Interrupt, backchannel, and pause data (\(\sim 36\)k h) drive the action channel to emit the right control labels at the right time, and to switch the assistant TA4 to silence within a small chunk-level latency under interruption.
    \item Tool-call data (\(\sim 14\)k h) drives the model to emit planning text plus structured tool calls on the action channel, both in standard turn-taking single- and multi-action requests and in topically unrelated backchannel-action requests that must not break the assistant's spoken thread.
\end{itemize}

\subsection{Loss with full-duplex-aware masking and reweighting}

The training objective is standard next-token cross-entropy on the assistant TA4 stream and the action channel, plus a general text-modelling term on the text-only data slice. On top of this base loss, we apply additional loss masks and per-token weights to selected state tokens and to specific positions in the chunked dual-stream three-channel sequence, so that the optimisation is better matched to the full-duplex training setting (e.g., silence anchors, channel-boundary markers, and task-conditioned segments are not trained as ordinary content tokens). The decoder interface itself does not change between stages or task families, which is what allows DuplexSLA to absorb very different supervision signals, such as duplex dialogue, dual-side ASR, timing control, and tool calls, within a single model.

\subsection{Stage division by capability}

A turn-based agent can be improved by adding more text or more tool examples. A duplex spoken agent carries the additional burden of \emph{timing}. The CPT stage therefore establishes the timing prior using ordinary duplex dialogue plus dual-side ASR, and the post-training stage sharpens it for pause, interrupt, backchannel, and tool calling. This division was the most data-efficient setup in our experiments: pure duplex dialogue alone teaches turn taking but not interaction control, while starting with capability-heavy data without first stabilizing the duplex serialization leads to noticeably worse speech smoothness on the assistant audio.

%% file: content/evaluation.tex
\section{Evaluation}
\label{sec:eval}

\subsection{Benchmark and evaluation protocol}
\label{sec:eval_protocol}

Existing duplex benchmarks measure pause, interruption, and backchannel behaviour~\citep{lin2025fullduplexbench}, but none of them jointly stress sub-second yielding under semantic interruption, backchannel detection inside the action channel, backchannel-triggered tool calling, and multi-action tool calling on a duplex timeline. We therefore evaluate DuplexSLA on \textbf{DuplexSLA-Bench}, a duplex benchmark with \(1{,}200\) turn-taking cases (300 each for \texttt{normal}, \texttt{pause}, \texttt{interrupt}, \texttt{backchannel}) plus a \(900\)-case tool-call subset (300 each for single-action, multi-action, and backchannel-action requests). All numbers below are reported on this benchmark.

For every turn-taking scenario we report two quantities: an accuracy (did the model behave correctly at all?) and a delay (how far the realised action time was from the desired semantic anchor, computed only on hits). The accuracy windows and delay anchors are summarised in \cref{tab:windows}.

\begin{table}[H]
\centering
\small
\setlength{\tabcolsep}{4pt}
\renewcommand{\arraystretch}{1.25}
\begin{tabular}{p{0.18\linewidth}p{0.46\linewidth}p{0.24\linewidth}}
\toprule
Scenario & Accuracy window & Delay definition \\
\midrule
\texttt{normal} & assistant speech onset \mbox{\(\in [t_{\text{ue}} - 0.2,\,+\infty)\)} & \(|t_{\text{speak}} - t_{\text{ue}}|\) \\
\texttt{pause} & same as \texttt{normal}, but on hesitation-rich user audio & \(|t_{\text{speak}} - t_{\text{ue}}|\) \\
\texttt{interrupt} & assistant stop time \mbox{\(\in [t_{\text{int}} - 1,\,t_{\text{int}} + 2]\)} & \(|t_{\text{stop}} - t_{\text{int}}|\) \\
\texttt{backchannel} & a stop-or-restart event \mbox{\(\in [t_{\text{bc-s}} - 0.2,\,t_{\text{bc-e}} + 2]\)} & \(|t_{\text{label}} - t_{\text{bc-e}}|\) \\
\bottomrule
\end{tabular}
\caption{Accuracy windows and delay definitions for the four turn-taking scenarios. \(t_{\text{ue}}\): user-end time; \(t_{\text{int}}\): semantic interrupt anchor; \(t_{\text{bc-s}}, t_{\text{bc-e}}\): start and end of the user backchannel utterance.}
\label{tab:windows}
\end{table}

Each test case is a duplex audio session with semantic anchor times annotated. The system under test is fed in chunked streaming mode, and the assistant audio output is post-processed by an external VAD to obtain speak and stop transitions; for DuplexSLA the action channel is also read directly, so that backchannel labels are evaluated even when there is no user-perceptible audio change. \cref{tab:duplex_eval} states the full procedure in set-theoretic form.

\begin{table}[H]
\centering
\small
\setlength{\tabcolsep}{6pt}
\renewcommand{\arraystretch}{1.35}
\begin{tabular}{@{}p{0.14\linewidth}p{0.80\linewidth}@{}}
\toprule
\textbf{Stage} & \textbf{Definition} \\
\midrule
Inputs & Streaming model \(\mathcal{S}\); test case \((\mathbf{u}, \mathcal{H}, s, \mathcal{A})\) with user audio \(\mathbf{u}\), dialogue history \(\mathcal{H}\), scenario \(s \in \{\texttt{normal}, \texttt{pause}, \texttt{interrupt}, \texttt{backchannel}\}\) and anchor set \(\mathcal{A}\); prefill flag \(p \in \{0, 1\}\) for the context-prefill protocol. \\
\midrule
1. Init & \(\mathcal{S} \leftarrow \textsc{reset}\).\newline If \(p = 1\), \(\mathcal{S} \leftarrow \textsc{prefill}(\mathcal{H})\).\newline Initialise event log \(E \leftarrow \varnothing\). \\
\midrule
2. Stream & Split \(\mathbf{u}\) into \(K\) chunks of \(160\,\text{ms}\). For \(k = 0,\ldots,K-1\): \(o_k = \mathcal{S}.\textsc{step}(u_k)\), \(t_k = 0.16\,k\). Append \((t_k, \texttt{speak}, \varnothing)\) when \(o_k\) emits assistant speech, else \((t_k, \texttt{stop}, \varnothing)\); and for each label \(\ell \in \textsc{labels}(o_k.\text{act})\) append \((t_k, \ell, o_k.\text{act})\) to \(E\). \\
\midrule
3. Score & Let \(W_s\) and anchor \(t^\star_s\) be given by \cref{tab:windows}, and let \(\tau_s = \texttt{speak}\) for \(s \in \{\texttt{normal}, \texttt{pause}\}\), \(\tau_s = \texttt{stop}\) for \(s = \texttt{interrupt}\), and \(\tau_s = \texttt{backchannel}\) for \(s = \texttt{backchannel}\). Define
\(\hat{e}_s = \arg\min\bigl\{\,t : (t, \tau_s, \cdot) \in E,\; t \in W_s\,\bigr\}\); then \(\textsc{acc} = \mathbb{1}[\hat{e}_s\text{ exists}]\) and \(\textsc{delay} = |\hat{e}_s.t - t^\star_s|\) whenever \(\textsc{acc} = 1\). For \(s = \texttt{backchannel}\), accuracy is relaxed to any \(\{\texttt{stop}, \texttt{speak}\}\) event inside \(W_s\), since closed-source baselines emit no backchannel label and \(\textsc{delay}\) is therefore reported only when one is present. \\
\bottomrule
\end{tabular}
\caption{Pseudocode in tabular, set-theoretic form for the duplex turn-taking evaluation on DuplexSLA-Bench. The three deterministic stages -- initialise, stream, score -- are shared across systems; only the streaming input and how action labels are read differ. Systems that cannot expose action labels fall back to audio-only events, which is why \(\textsc{delay}\) on \texttt{backchannel} is N/A for non-DuplexSLA baselines (\cref{tab:fd_prefill}).}
\label{tab:duplex_eval}
\end{table}

\subsection{Tool-call results}

Each tool-call test case is a context plus a user request that requires one or more functions. Both systems are run under the same streaming protocol: DuplexSLA reads tool calls directly off the action channel, while the ASR + LLM cascade emits them in a post-ASR planning step. A predicted tool call counts as correct when (1) every ground-truth action has a predicted action with the same function name; (2) the arguments match -- exact match, both empty, or judged semantically consistent by an LLM with no ``core information conflict''; and (3) the trigger time is legal -- not earlier than the ground-truth offset by more than \(1.0\) s, and not later than the end of the audio by more than \(3.0\) s. Accuracy is the fraction of cases in which all ground-truth actions are matched, and delay is the average gap on matched actions.

\cref{tab:toolcall} reports accuracy and delay across the three call patterns. DuplexSLA matches the cascade in accuracy while delivering \(\sim 4\)x lower tool-call delay on average.

\begin{table}[H]
\centering
\footnotesize
\setlength{\tabcolsep}{4pt}
\begin{tabular}{lcccccccc}
\toprule
\multirow{2}{*}{Model} & \multicolumn{2}{c}{Avg.\,(3 patterns)} & \multicolumn{2}{c}{Single action} & \multicolumn{2}{c}{Multi actions} & \multicolumn{2}{c}{Backchannel action} \\
\cmidrule(lr){2-3}\cmidrule(lr){4-5}\cmidrule(lr){6-7}\cmidrule(lr){8-9}
& Acc.\,(\%) & Delay (s) & Acc.\,(\%) & Delay (s) & Acc.\,(\%) & Delay (s) & Acc.\,(\%) & Delay (s) \\
\midrule
ASR + LLM sys & 91.33 & 2.77 & 89.33 & 2.33 & 89.33 & 4.71 & 95.33 & 1.27 \\
\textbf{DuplexSLA} & 85.56 & \textbf{0.64} & 85.67 & \textbf{0.67} & 75.00 & \textbf{0.68} & 96.00 & \textbf{0.57} \\
\bottomrule
\end{tabular}
\caption{Tool-call results on the DuplexSLA-Bench tool-call subset (\(900\) cases). The cascade baseline replaces DuplexSLA with a streaming ASR module whose transcript is fed to an LLM that emits tool calls. DuplexSLA is competitive in accuracy and \(\sim 4\)x faster in tool-call delay on average.}
\label{tab:toolcall}
\end{table}

\subsection{Full-duplex turn-taking results}

We evaluate full-duplex behaviour in two settings that mirror typical deployments.

\paragraph{(1) Context prefill.}
Systems that can prefill the entire dialogue history are evaluated on all four scenarios -- \texttt{normal}, \texttt{pause}, \texttt{interrupt}, \texttt{backchannel} -- alongside DuplexSLA. \cref{tab:fd_prefill} reports the results. DuplexSLA is the only system that handles backchannel correctly (\(98.33\%\) vs.\ \(\le 40\%\) for all baselines), and it achieves the lowest delay in every scenario. The backchannel delay column is \emph{N/A} for the closed-source baselines because they expose no backchannel label, so the only audio-derived signal is the absence of an interruption.

\begin{table}[H]
\centering
\scriptsize
\setlength{\tabcolsep}{4pt}
\begin{tabular}{lcccccccc}
\toprule
\multirow{2}{*}{Model} & \multicolumn{2}{c}{\texttt{normal}} & \multicolumn{2}{c}{\texttt{pause}} & \multicolumn{2}{c}{\texttt{interrupt}} & \multicolumn{2}{c}{\texttt{backchannel}} \\
\cmidrule(lr){2-3}\cmidrule(lr){4-5}\cmidrule(lr){6-7}\cmidrule(lr){8-9}
& Acc.\,(\%) & Delay (s) & Acc.\,(\%) & Delay (s) & Acc.\,(\%) & Delay (s) & Acc.\,(\%) & Delay (s) \\
\midrule
\textbf{DuplexSLA} & \textbf{96.00} & \textbf{0.27} & \textbf{93.33} & \textbf{0.27} & \textbf{99.33} & \textbf{0.40} & \textbf{98.33} & \textbf{0.32} \\
gemini-3.1-flash-live & 93.67 & 1.18 & 94.33 & 1.17 & 63.67 & 0.62 & 40.00 & N/A \\
gpt-realtime-1.5 (semantic-vad-high) & 91.33 & 1.67 & 90.33 & 1.68 & 79.00 & 0.68 & 0.33 & N/A \\
gpt-realtime-1.5 (server-vad-40ms) & 82.33 & 0.95 & 71.00 & 1.02 & 77.00 & 0.72 & 13.00 & N/A \\
\bottomrule
\end{tabular}
\caption{Full-duplex turn-taking results in the context-prefill setting. DuplexSLA achieves the highest accuracy and lowest delay in all four scenarios. Closed-source baselines do not expose a backchannel label, so their backchannel delay is N/A.}
\label{tab:fd_prefill}
\end{table}

\paragraph{(2) No context prefill.}
Many duplex systems cannot cheaply preload long histories, so we also evaluate a no-prefill setting that reduces to the two scenarios all systems support: \texttt{normal} and \texttt{pause}. \cref{tab:fd_no_prefill} reports accuracy and delay against open-source duplex backbones~\citep{wang2024freezeomni,roy2026personaplex} and commercial APIs. DuplexSLA again achieves the lowest delay (\(\sim 0.30\) s) while staying among the highest-accuracy systems (\(94.34\%\) average), and is the only sub-second model with competitive accuracy.

\begin{table}[H]
\centering
\scriptsize
\setlength{\tabcolsep}{4pt}
\begin{tabular}{lcccccc}
\toprule
\multirow{2}{*}{Model} & \multicolumn{2}{c}{Avg.\,(2 scenarios)} & \multicolumn{2}{c}{\texttt{normal}} & \multicolumn{2}{c}{\texttt{pause}} \\
\cmidrule(lr){2-3}\cmidrule(lr){4-5}\cmidrule(lr){6-7}
& Acc.\,(\%) & Delay (s) & Acc.\,(\%) & Delay (s) & Acc.\,(\%) & Delay (s) \\
\midrule
\textbf{DuplexSLA} & \textbf{94.34} & \textbf{0.30} & \textbf{95.67} & \textbf{0.29} & \textbf{93.00} & \textbf{0.31} \\
Freeze-Omni & 10.67 & 0.36 & 10.33 & 0.40 & 11.00 & 0.33 \\
PersonaPlex & 22.34 & 0.47 & 22.67 & 0.38 & 22.00 & 0.55 \\
MiniCPM-o & 82.00 & 0.61 & 83.33 & 0.62 & 80.67 & 0.59 \\
gemini-3.1-flash-live & 93.17 & 1.17 & 93.67 & 1.16 & 93.67 & 1.18 \\
gpt-realtime-1.5 (semantic-vad-high) & 96.50 & 1.57 & 96.70 & 1.57 & 96.30 & 1.57 \\
gpt-realtime-1.5 (server-vad-40ms) & 85.50 & 0.83 & 91.30 & 0.83 & 79.70 & 0.83 \\
\bottomrule
\end{tabular}
\caption{Full-duplex turn-taking results in the no-context-prefill setting. DuplexSLA is the only sub-second system with competitive accuracy; commercial APIs reach high accuracy but pay \(>\!1\) s in delay. Open-source duplex backbones without targeted post-training collapse on the \texttt{pause} subset, illustrating that pause robustness has to be supervised explicitly.}
\label{tab:fd_no_prefill}
\end{table}

\subsection{Take-aways}

Two patterns are consistent across the tables. (1) On turn taking, DuplexSLA delivers sub-second responses in all four scenarios and is the only system that cleanly handles backchannel detection. (2) On tool calling, DuplexSLA matches the cascade in accuracy at \(\sim 4\)x lower delay, because the action channel emits planning and tool calls without waiting for a turn boundary or interrupting assistant audio. Together they validate the central design choice -- an explicit action channel on top of a duplex backbone, supervised by the data recipe in \cref{sec:data}.

%% file: content/conclusion.tex
\section{Conclusion}

In this work, we propose DuplexSLA, a native full-duplex \emph{speech-language-action} foundation model that places listening, speaking, planning, and tool calling on a shared \(160\) ms chunk timeline. The dual-stream three-channel formulation -- continuous user audio, discrete assistant TA4, and a rate-limited textual action channel -- lets a single backbone learn (1) semantic-driven interruption, pause, and backchannel without an external semantic VAD, and (2) in-conversation planning and tool calling that runs in parallel with assistant speech. Trained on a duplex-dialogue plus dual-side ASR pretraining mixture and a smaller capability-oriented post-training mixture, DuplexSLA achieves sub-second latency in all four turn-taking scenarios on the proposed DuplexSLA-Bench, while remaining competitive with an ASR + LLM cascade on tool-call accuracy at \(3\!-\!4\)x lower latency. We view DuplexSLA as a step toward duplex spoken agents that combine fluent speech with timely action, and we expect the action-channel design to extend naturally to richer planning signals, multi-turn agentic workflows, and broader open-domain spoken tool use.

%% file: content/Contributors.tex
% Contributors section intentionally left blank in this draft.

%% file: content/appendix.tex
\section{Per-Chunk Serialization Case Studies}
\label{app:serialization}

This appendix gives concrete chunk-by-chunk traces of the dual-stream three-channel format introduced in \cref{sec:arch}. For readability, each row corresponds to one chunk (\(160\) ms), the user audio segment is abbreviated as ``\(U\,U\)'', and the TA4 unit is shown as ``\(T(\cdot)\,A\,A\,A\,A\)'', where \(T\) holds either a text token (e.g., ``\zh{确}''), the silence anchor \texttt{<vad\_silence>}, or the trailing pad \texttt{<tts\_pad>}. The action segment, when not empty, is shown after the assistant TA4 and before the chunk-terminating \texttt{<|action\_end|>}.

\subsection{User-channel ASR}
The user-channel transcript is emitted on the action channel with a fixed lag of \(2\) chunks (\(320\) ms). Tokens that fall in the same chunk are merged.
\begin{tcolorbox}[colback=casebg, colframe=black, boxrule=0.6pt, arc=2mm, breakable, fontupper=\footnotesize\ttfamily]
\setlength{\parskip}{0pt}\setlength{\parindent}{0pt}\raggedright
<|user\_audio\_begin|> U U(\zh{今天}) <|user\_audio\_end|> <|assistant\_audio\_begin|> T(<vad\_silence>) A A A A <|assistant\_audio\_end|> <|action\_end|>\\
<|user\_audio\_begin|> U U(\zh{天气很}) <|user\_audio\_end|> <|assistant\_audio\_begin|> T(<vad\_silence>) A A A A <|assistant\_audio\_end|> <|action\_end|>\\
<|user\_audio\_begin|> U U(\zh{好}) <|user\_audio\_end|> <|assistant\_audio\_begin|> T(<vad\_silence>) A A A A <|assistant\_audio\_end|> \zh{今天} <|action\_end|>\\
<|user\_audio\_begin|> U U <|user\_audio\_end|> <|assistant\_audio\_begin|> T(<vad\_silence>) A A A A <|assistant\_audio\_end|> \zh{天气很} <|action\_end|>\\
<|user\_audio\_begin|> U U <|user\_audio\_end|> <|assistant\_audio\_begin|> T(\zh{确}) A A A A <|assistant\_audio\_end|> \zh{好} <|action\_end|>\\
<|user\_audio\_begin|> U U <|user\_audio\_end|> <|assistant\_audio\_begin|> T(\zh{实}) A A A A <|assistant\_audio\_end|> <|action\_end|>\\
<|user\_audio\_begin|> U U <|user\_audio\_end|> <|assistant\_audio\_begin|> T(<tts\_pad>) A A A A <|assistant\_audio\_end|> <|action\_end|>\\
<|user\_audio\_begin|> U U <|user\_audio\_end|> <|assistant\_audio\_begin|> T(<vad\_silence>) A A A A <|assistant\_audio\_end|> <|action\_end|>\\
<|user\_audio\_begin|> U U <|user\_audio\_end|> <|assistant\_audio\_begin|> T(<vad\_silence>) A A A A <|assistant\_audio\_end|> <|action\_end|>
\end{tcolorbox}

\subsection{Assistant-channel ASR}

\paragraph{Necessity of assistant-side ASR.}
The action channel cannot simply delay-copy the \(T\) tokens of the assistant TA4 by two chunks. The TA4 layout looks like
\[
T A_4\;\; T A_4\;\; T A_4\;\; T A_4\;\; T(\text{\texttt{<tts\_pad>}}) A_4 \;\;T(\text{\texttt{<tts\_pad>}}) A_4 \dots
\]
That is, the assistant text tokens are \emph{left-aligned} inside the chunk and do not carry exact timing: a single Chinese word can be packed into the first \(T\) slot of a chunk while the corresponding audio is actually played in the next chunk. Re-emitting the assistant transcript on the action channel at the chunk where each character is genuinely being spoken therefore teaches DuplexSLA the correct time alignment between assistant audio and action time.

\begin{tcolorbox}[colback=casebg, colframe=black, boxrule=0.6pt, arc=2mm, breakable, fontupper=\footnotesize\ttfamily]
\setlength{\parskip}{0pt}\setlength{\parindent}{0pt}\raggedright
<|user\_audio\_begin|> U U(\zh{今天}) <|user\_audio\_end|> <|assistant\_audio\_begin|> T(<vad\_silence>) A A A A <|assistant\_audio\_end|> <|action\_end|>\\
<|user\_audio\_begin|> U U(\zh{天气很}) <|user\_audio\_end|> <|assistant\_audio\_begin|> T(<vad\_silence>) A A A A <|assistant\_audio\_end|> <|action\_end|>\\
<|user\_audio\_begin|> U U(\zh{好}) <|user\_audio\_end|> <|assistant\_audio\_begin|> T(<vad\_silence>) A A A A <|assistant\_audio\_end|> <|action\_end|>\\
<|user\_audio\_begin|> U U <|user\_audio\_end|> <|assistant\_audio\_begin|> T(<vad\_silence>) A A A A <|assistant\_audio\_end|> <|action\_end|>\\
<|user\_audio\_begin|> U U <|user\_audio\_end|> <|assistant\_audio\_begin|> T(\zh{确}) A A A A <|assistant\_audio\_end|> <|action\_end|>\\
<|user\_audio\_begin|> U U <|user\_audio\_end|> <|assistant\_audio\_begin|> T(\zh{实}) A A A A <|assistant\_audio\_end|> <|action\_end|>\\
<|user\_audio\_begin|> U U <|user\_audio\_end|> <|assistant\_audio\_begin|> T(<tts\_pad>) A A A A <|assistant\_audio\_end|> \zh{确} <|action\_end|>\\
<|user\_audio\_begin|> U U <|user\_audio\_end|> <|assistant\_audio\_begin|> T(<vad\_silence>) A A A A <|assistant\_audio\_end|> <|action\_end|>\\
<|user\_audio\_begin|> U U <|user\_audio\_end|> <|assistant\_audio\_begin|> T(<vad\_silence>) A A A A <|assistant\_audio\_end|> \zh{实} <|action\_end|>
\end{tcolorbox}

\subsection{Duplex dialogue with planning + tool call}

The user comments that the car is cold; once the request becomes semantically clear, the action channel emits a planning fragment and a tool call (\texttt{set\_car\_setting}). The assistant audio remains coherent across the same chunks. Because of the per-chunk action-token cap (\(\le 10\) tokens), the planning text and the tool-call body span several consecutive chunks rather than a single one, following the FIFO rule in \cref{sec:fifo}. The strict time-alignment prior installed by dual-side ASR during CPT (\cref{sec:training}) lets the model treat this multi-chunk spillover as a bounded, budget-induced transmission delay rather than as drift in the offset itself: the first action token still emerges at the chunk that owns the semantic anchor, so the realised trigger time remains aligned with the annotated offset.

\begin{tcolorbox}[colback=casebg, colframe=black, boxrule=0.6pt, arc=2mm, breakable, fontupper=\footnotesize\ttfamily]
\setlength{\parskip}{0pt}\setlength{\parindent}{0pt}\raggedright
<|user\_audio\_begin|> U U(\zh{车里}) <|user\_audio\_end|> <|assistant\_audio\_begin|> T(<vad\_silence>) A A A A <|assistant\_audio\_end|> <|action\_end|>\\
<|user\_audio\_begin|> U U(\zh{真冷}) <|user\_audio\_end|> <|assistant\_audio\_begin|> T(<vad\_silence>) A A A A <|assistant\_audio\_end|> <|action\_end|>\\
<|user\_audio\_begin|> U U(\zh{啊}) <|user\_audio\_end|> <|assistant\_audio\_begin|> T(<vad\_silence>) A A A A <|assistant\_audio\_end|> \zh{用户感到有点冷，应该打开空调开暖} <|action\_end|>\\
<|user\_audio\_begin|> U U <|user\_audio\_end|> <|assistant\_audio\_begin|> T(<vad\_silence>) A A A A <|assistant\_audio\_end|> \zh{风，调整到合适的温度}<|toolcall\_begin|>\{"function":\,"set\_car\_setting", <|action\_end|>\\
<|user\_audio\_begin|> U U <|user\_audio\_end|> <|assistant\_audio\_begin|> T(\zh{我}) A A A A <|assistant\_audio\_end|> "arguments":\,"AC=26\zh{度}"\}<|toolcall\_end|> <|action\_end|>\\
<|user\_audio\_begin|> U U <|user\_audio\_end|> <|assistant\_audio\_begin|> T(\zh{明白}) A A A A <|assistant\_audio\_end|> <|action\_end|>\\
<|user\_audio\_begin|> U U <|user\_audio\_end|> <|assistant\_audio\_begin|> T(\zh{了}) A A A A <|assistant\_audio\_end|> <|action\_end|>\\
<|user\_audio\_begin|> U U <|user\_audio\_end|> <|assistant\_audio\_begin|> T(\zh{这就}) A A A A <|assistant\_audio\_end|> <|action\_end|>\\
<|user\_audio\_begin|> U U <|user\_audio\_end|> <|assistant\_audio\_begin|> T(\zh{帮}) A A A A <|assistant\_audio\_end|> <|action\_end|>\\
<|user\_audio\_begin|> U U <|user\_audio\_end|> <|assistant\_audio\_begin|> T(\zh{你}) A A A A <|assistant\_audio\_end|> <|action\_end|>\\
<|user\_audio\_begin|> U U <|user\_audio\_end|> <|assistant\_audio\_begin|> T(\zh{打开}) A A A A <|assistant\_audio\_end|> <|action\_end|>\\
<|user\_audio\_begin|> U U <|user\_audio\_end|> <|assistant\_audio\_begin|> T(\zh{空调}) A A A A <|assistant\_audio\_end|> <|action\_end|>\\
<|user\_audio\_begin|> U U <|user\_audio\_end|> <|assistant\_audio\_begin|> T(<tts\_pad>) A A A A <|assistant\_audio\_end|> <|action\_end|>\\
<|user\_audio\_begin|> U U <|user\_audio\_end|> <|assistant\_audio\_begin|> T(<tts\_pad>) A A A A <|assistant\_audio\_end|> <|action\_end|>\\
<|user\_audio\_begin|> U U <|user\_audio\_end|> <|assistant\_audio\_begin|> T(<tts\_pad>) A A A A <|assistant\_audio\_end|> <|action\_end|>\\
<|user\_audio\_begin|> U U <|user\_audio\_end|> <|assistant\_audio\_begin|> T(<vad\_silence>) A A A A <|assistant\_audio\_end|> <|action\_end|>
\end{tcolorbox}

\subsection{An illustrative interrupt case}
\label{app:interrupt}

The user starts to follow up on the assistant's previous answer (``you are right'') and then immediately rolls into a real new statement (``but the project schedule is tight, I don't really have a choice''). DuplexSLA is expected to keep speaking while the user is only acknowledging, and then yield within a small chunk-level latency once the new content starts. The action channel emits an \texttt{interrupt} label, and the assistant TA4 immediately switches from text and audio tokens to silence anchors. This trace mirrors the right-hand panel of \cref{fig:interrupt_backchannel}.

\begin{tcolorbox}[colback=casebg, colframe=black, boxrule=0.6pt, arc=2mm, breakable, fontupper=\footnotesize\ttfamily]
\setlength{\parskip}{0pt}\setlength{\parindent}{0pt}\raggedright
<|user\_audio\_begin|> U U(\zh{你说}) <|user\_audio\_end|> <|assistant\_audio\_begin|> T(\zh{身体}) A A A A <|assistant\_audio\_end|> <|action\_end|>\\
<|user\_audio\_begin|> U U(\zh{得对}) <|user\_audio\_end|> <|assistant\_audio\_begin|> T(\zh{健康}) A A A A <|assistant\_audio\_end|> <|action\_end|>\\
<|user\_audio\_begin|> U U <|user\_audio\_end|> <|assistant\_audio\_begin|> T(\zh{是}) A A A A <|assistant\_audio\_end|> <|action\_end|>\\
<|user\_audio\_begin|> U U(\zh{但项}) <|user\_audio\_end|> <|assistant\_audio\_begin|> T(\zh{最}) A A A A <|assistant\_audio\_end|> <|action\_end|>\\
<|user\_audio\_begin|> U U(\zh{目很}) <|user\_audio\_end|> <|assistant\_audio\_begin|> T(\zh{重要}) A A A A <|assistant\_audio\_end|> <|action\_end|>\\
<|user\_audio\_begin|> U U(\zh{紧}) <|user\_audio\_end|> <|assistant\_audio\_begin|> T(\zh{的}) A A A A <|assistant\_audio\_end|> \zh{检测到用户插话}<interrupt> <|action\_end|>\\
<|user\_audio\_begin|> U U(\zh{我也}) <|user\_audio\_end|> <|assistant\_audio\_begin|> T(<vad\_silence>) A A A A <|assistant\_audio\_end|> <|action\_end|>\\
<|user\_audio\_begin|> U U(\zh{没办}) <|user\_audio\_end|> <|assistant\_audio\_begin|> T(<vad\_silence>) A A A A <|assistant\_audio\_end|> <|action\_end|>\\
<|user\_audio\_begin|> U U(\zh{法}) <|user\_audio\_end|> <|assistant\_audio\_begin|> T(<vad\_silence>) A A A A <|assistant\_audio\_end|> <|action\_end|>
\end{tcolorbox}

The example also shows that the same acknowledgement (``you are right''), in the backchannel scenario, would be followed by user silence rather than a new statement; in that case DuplexSLA emits a \texttt{backchannel} label on the action channel and continues speaking.

\subsection{An illustrative backchannel case}
\label{app:backchannel}

In this case, the assistant is in the middle of a long answer; the user inserts a brief ``\zh{没错}\,/\,that's right'' acknowledgement and then stays silent. The expected behaviour is that the action channel emits a \texttt{backchannel} label inside the user-utterance window, while the assistant TA4 keeps producing the planned response. This mirrors the left-hand panel of \cref{fig:interrupt_backchannel}.

\begin{tcolorbox}[colback=casebg, colframe=black, boxrule=0.6pt, arc=2mm, breakable, fontupper=\footnotesize\ttfamily]
\setlength{\parskip}{0pt}\setlength{\parindent}{0pt}\raggedright
<|user\_audio\_begin|> U U <|user\_audio\_end|> <|assistant\_audio\_begin|> T(\zh{相依为}) A A A A <|assistant\_audio\_end|> <|action\_end|>\\
<|user\_audio\_begin|> U U(\zh{没}) <|user\_audio\_end|> <|assistant\_audio\_begin|> T(\zh{命的}) A A A A <|assistant\_audio\_end|> <|action\_end|>\\
<|user\_audio\_begin|> U U(\zh{错}) <|user\_audio\_end|> <|assistant\_audio\_begin|> T(\zh{感觉}) A A A A <|assistant\_audio\_end|> \zh{检测到附和语气}<backchannel> <|action\_end|>\\
<|user\_audio\_begin|> U U <|user\_audio\_end|> <|assistant\_audio\_begin|> T(\zh{比}) A A A A <|assistant\_audio\_end|> <|action\_end|>\\
<|user\_audio\_begin|> U U <|user\_audio\_end|> <|assistant\_audio\_begin|> T(\zh{直接}) A A A A <|assistant\_audio\_end|> <|action\_end|>\\
<|user\_audio\_begin|> U U <|user\_audio\_end|> <|assistant\_audio\_begin|> T(\zh{撒糖}) A A A A <|assistant\_audio\_end|> <|action\_end|>\\
<|user\_audio\_begin|> U U <|user\_audio\_end|> <|assistant\_audio\_begin|> T(\zh{有意思}) A A A A <|assistant\_audio\_end|> <|action\_end|>\\
<|user\_audio\_begin|> U U <|user\_audio\_end|> <|assistant\_audio\_begin|> T(\zh{多了}) A A A A <|assistant\_audio\_end|> <|action\_end|>
\end{tcolorbox}

\section{Action vocabulary and task-conditioned system prompts}
\label{app:vocab}

Beyond textual planning and tool-call JSON, the action channel uses a small set of canonical control-label phrases. During data construction, the same \texttt{name} field is paraphrased by several near-synonyms so that the model is not over-fit to a single surface form. \cref{tab:action_vocab} lists the canonical name and representative paraphrases observed in the training data, and \cref{tab:sysprompts} summarises the per-task system prompt used during training, where the assistant-speaker placeholder \(\{\cdot\}\) is filled with one of the \(18\) main voice-clone speakers.

\begin{table}[H]
\centering
\small
\begin{tabular}{p{0.15\linewidth}p{0.18\linewidth}p{0.60\linewidth}}
\toprule
Action \texttt{name} & Trigger context & Representative canonical phrases (Chinese, with paraphrases) \\
\midrule
\texttt{response} & user finishes a turn & \zh{用户发言结束} / \zh{检测到表达完毕} / \zh{接收到完整内容} \\
\texttt{interrupt} & user starts a real new thought during assistant speech & \zh{检测到用户插话} / \zh{识别到插话意图} / \zh{检测到有效发言} \\
\texttt{backchannel} & user emits short feedback without taking the floor & \zh{检测到附和语气} / \zh{识别到轻微反馈} / \zh{用户仅做确认} \\
\texttt{asr} & duplex ASR supervision (Appendix~A.1, A.2) & no canonical phrase; planning text is the delayed transcript token \\
tool name & tool-use scenario & free planning text plus structured JSON: \newline {\scriptsize\texttt{<|toolcall\_begin|>\{"function":\,\(f\),\,"arguments":\,\(\theta\)\}<|toolcall\_end|>}} \\
\bottomrule
\end{tabular}
\caption{Canonical control labels used on the action channel. The label set is kept compact so that turn-taking decisions are decoupled from spoken content.}
\label{tab:action_vocab}
\end{table}

\begin{table}[H]
\centering
\small
\begin{tabular}{p{0.20\linewidth}p{0.72\linewidth}}
\toprule
Task family & Training-time system prompt \\
\midrule
\texttt{dialogue} & (empty) \\
\texttt{asr\_human} & \zh{请记录下你所听到的语音内容，只记录用户说的内容。} \\
\texttt{asr\_assistant} & \zh{请记录下你所听到的语音内容，只记录助手说的内容。} \\
\texttt{interpret} & \zh{请翻译用户说的内容。} \\
\texttt{toolcall} & \zh{你是一个专注于与人互动的 AI，既能聊天，也能使用工具来解决用户的问题。} \\
\texttt{interrupt}    & \zh{你是一个 AI 语音助手，用 \(\{\,\cdot\,\}\) 的声音来说话。} \\
\texttt{backchannel}  & \zh{你是一个 AI 语音助手，用 \(\{\,\cdot\,\}\) 的声音来说话。} \\
\texttt{pause}        & \zh{你是一个 AI 语音助手，用 \(\{\,\cdot\,\}\) 的声音来说话。} \\
\bottomrule
\end{tabular}
\caption{Per-task system prompt used during training. The model picks up the task signal from the system prompt; \(\{\,\cdot\,\}\) is filled at sample-build time with the name of one of the \(18\) canonical assistant speakers.}
\label{tab:sysprompts}
\end{table}

\section{DuplexSLA-Bench composition and tool schema coverage}
\label{app:bench}

DuplexSLA-Bench is a benchmark of \(2{,}100\) cases used in \cref{sec:eval}. It is organised into two subsets that exercise the two highlight capabilities of DuplexSLA. \cref{tab:bench_composition} summarises the case counts, and \cref{tab:tool_schema} lists the full set of \(50\) tool schemas exercised by the tool-call subset, grouped by intent family.

\begin{table}[H]
\centering
\small
\begin{tabular}{p{0.32\linewidth}rp{0.46\linewidth}}
\toprule
Subset & \#cases & Notes \\
\midrule
\multicolumn{3}{l}{\emph{Turn-taking subset (\(1{,}200\) cases, \cref{tab:fd_prefill,tab:fd_no_prefill}).}} \\
\quad \texttt{normal} & \(300\) & ordinary end-of-turn response \\
\quad \texttt{pause} & \(300\) & hesitation-rich within-turn silence \\
\quad \texttt{interrupt} & \(300\) & semantic interruption mid-assistant-speech \\
\quad \texttt{backchannel} & \(300\) & short user feedback without floor transfer \\
\midrule
\multicolumn{3}{l}{\emph{Tool-call subset (\(900\) cases, \cref{tab:toolcall}).}} \\
\quad \texttt{single-action} & \(300\) & one explicit user request, single function \\
\quad \texttt{multi-action} & \(300\) & one user turn, multiple ordered functions \\
\quad \texttt{backchannel-action} & \(300\) & topically unrelated function triggered while assistant keeps speaking \\
\bottomrule
\end{tabular}
\caption{DuplexSLA-Bench composition.}
\label{tab:bench_composition}
\end{table}

\renewcommand{\arraystretch}{1.1}
\begin{longtable}{@{}p{0.32\linewidth} p{0.62\linewidth}@{}}
\caption{Full cabin and smart-home tool schema used during training and evaluation. The schema contains \(50\) functions, organised into car-cabin control, navigation, media, and broader on-device search and query intents. Backchannel, interrupt, and pause are \emph{not} part of this schema -- they are handled by the dedicated control-label vocabulary in \cref{tab:action_vocab}.}
\label{tab:tool_schema} \\
\toprule
Function name & Description \\
\midrule
\endfirsthead
\multicolumn{2}{l}{\emph{(continued from previous page)}} \\
\toprule
Function name & Description \\
\midrule
\endhead
\midrule
\multicolumn{2}{r}{\emph{(continued on next page)}} \\
\endfoot
\bottomrule
\endlastfoot

\multicolumn{2}{@{}l}{\emph{Cabin and hardware control.}} \\
\texttt{open\_car\_setting}        & Turn on a car-related hardware feature or software setting (AC, windows, defrost, etc.). \\
\texttt{close\_car\_setting}       & Turn off a car-related hardware feature or software setting. \\
\texttt{set\_car\_setting}         & Set a car-related setting to a target value (AC temperature, seat position, etc.). \\
\texttt{increase\_car\_setting}    & Increase a car-related setting value (raise AC temperature, raise volume, etc.). \\
\texttt{decrease\_car\_setting}    & Decrease a car-related setting value. \\
\texttt{query\_car\_setting}       & Query the current state of a car setting (AC temperature, seat position, etc.). \\
\texttt{save\_car\_setting}        & Save the current car settings for quick recall later. \\
\texttt{set\_pet\_car\_setting}    & Configure pet-aware cabin settings (pet mode, climate). \\
\texttt{set\_car\_alarm}           & Configure the car alarm system (alarm time, on or off). \\

\midrule
\multicolumn{2}{@{}l}{\emph{System-level settings and apps.}} \\
\texttt{open\_system\_setting}        & Open a system-level settings panel (display, sound, etc.). \\
\texttt{close\_system\_setting}       & Close a system-level settings panel. \\
\texttt{set\_system\_setting}         & Set a system-level value (display brightness, master volume, etc.). \\
\texttt{increase\_system\_setting}    & Increase a system-level value (volume, brightness). \\
\texttt{decrease\_system\_setting}    & Decrease a system-level value (volume, brightness). \\
\texttt{disconnect\_system\_setting}  & Disconnect a system-level connection (Bluetooth, Wi-Fi). \\
\texttt{open\_app}                    & Open an in-vehicle application. \\
\texttt{close\_app}                   & Close an in-vehicle application. \\
\texttt{switch\_page}                 & Switch between UI pages via voice. \\
\texttt{scroll}                       & Scroll the UI vertically or horizontally. \\
\texttt{select\_option}               & Select one option from a multi-choice prompt. \\

\midrule
\multicolumn{2}{@{}l}{\emph{Navigation.}} \\
\texttt{navigate}                  & Start route planning and turn-by-turn navigation. \\
\texttt{change\_navigation\_route} & Change the current navigation route. \\
\texttt{add\_waypoint}             & Add a waypoint to the current route. \\
\texttt{remove\_waypoint}          & Remove a waypoint from the current route. \\
\texttt{resume\_navigation}        & Resume the active navigation session. \\
\texttt{query\_arrival\_time}      & Query the estimated arrival time at the destination. \\
\texttt{query\_distance}           & Query the distance between two locations. \\
\texttt{query\_road\_conditions}   & Query current road and traffic conditions. \\
\texttt{search\_along\_route}      & Search points of interest along the planned route (gas, restaurants, parking). \\

\midrule
\multicolumn{2}{@{}l}{\emph{Media playback.}} \\
\texttt{play\_media}        & Play a piece of media content (music, podcast, video). \\
\texttt{play\_broadcast}    & Play radio or broadcast content (radio shows, news). \\
\texttt{play\_online\_video}& Play an online video. \\
\texttt{search\_music}      & Search the music library for songs, albums or artists. \\
\texttt{search\_online\_video} & Search online video platforms for a specific clip. \\
\texttt{next\_track}        & Advance to the next track or media item. \\
\texttt{previous\_track}    & Go back to the previous track or media item. \\

\midrule
\multicolumn{2}{@{}l}{\emph{Search and queries.}} \\
\texttt{search\_food}          & Search for food-related information (restaurants, dishes). \\
\texttt{search\_entertainment} & Search for entertainment content (movies, music, etc.). \\
\texttt{search\_lifestyle}     & Search for lifestyle services and information (housekeeping, gyms, events). \\
\texttt{search\_shopping}      & Search for shopping information (items, stores, deals). \\
\texttt{search\_scenic\_spot}  & Search for tourist attractions (intro, hours, ticket price). \\
\texttt{search\_hotel}         & Search hotel information (name, location, booking). \\
\texttt{search\_travel}        & Search travel options (flight, train, public transit). \\
\texttt{search\_building}      & Search information about buildings (name, location, history). \\
\texttt{search\_recognition}   & Run a voice-recognition-driven search to identify and look up content. \\
\texttt{query\_weather}        & Query current or forecast weather. \\
\texttt{query\_calendar}       & Query calendar information (dates, holidays, events). \\
\texttt{query\_stock}          & Query stock-market information (price, trend). \\
\texttt{generate\_text\_resource} & Turn text into a resource (document, report, etc.). \\
\texttt{make\_call}            & Place a phone call to a contact or to a dialed number. \\
\end{longtable}

\section{Inference and serving notes}
\label{app:serving}

The decisions that matter for serving are summarised below; the per-chunk action-token cap is a deployment budget rather than an architectural constant (\cref{sec:budget}).

\begin{itemize}
    \item Conversational clock: \(\Delta = 160\) ms.
    \item Per-chunk model output: \(5\) assistant TA4 tokens (always) plus up to \(10\) action text tokens.
    \item User audio is encoded by a causal speech front end; no future user audio is required to advance one chunk.
    \item When the action stream needs to emit more than \(10\) tokens in one chunk (e.g., a long planning text plus a JSON tool-call body), the surplus spills into the action segment of the next chunk. As a result, tool-call closing markers can land in a later chunk than the opening marker, but the trigger time of the action object is always anchored to the chunk where the planning text starts.
\end{itemize}

\section{Action object schema}
\label{app:schema}

For completeness, each action object on the action channel carries the following conceptual fields. Concrete storage layouts vary across data families, but the abstract schema below is shared by all of them.

\begin{itemize}
    \item \texttt{name}: function name -- one of the \(50\) cabin and smart-home tool schemas, or one of \texttt{response}, \texttt{interrupt}, \texttt{backchannel}, or \texttt{asr}.
    \item \texttt{planning}: optional natural-language planning text, kept short so that it fits within a few chunks.
    \item \texttt{parameters}: optional JSON-style argument dictionary; for \texttt{asr} or control-only labels this is empty.
    \item \texttt{offset}: semantic trigger time on the conversational clock, snapped to a chunk index at training time.
\end{itemize}
The action stream is a single FIFO queue keyed by \texttt{offset} (\cref{sec:fifo}): when two or more action objects fall into the same chunk, they are serialized in trigger-time order on the action channel (ties broken by id); when an action's tokens exceed the per-chunk \(\le\!10\)-token budget, the surplus spills into the following chunks while any later-triggered action waits in the queue. This is what allows multi-action tool calls (\cref{fig:toolcall_capabilities}b) to share a short timeline window without losing temporal interpretability or breaking the assistant audio.